# A note on geodesics in inhomogeneous expanding spacetimes

**D Pérez**[1], **G E Romero**[1,2], **L Combi**[1,3], **E Gutiérrez**[1]

[1] Instituto Argentino de Radioastronomía (CCT-La Plata, CONICET; CICPBA), C.C. No. 5, 1894, Villa Elisa, Argentina

[2] Facultad de Ciencias Astronómicas y Geofísicas, Universidad Nacional de La Plata, Paseo del Bosque s/n, 1900 La Plata, Buenos Aires, Argentina

[3] Departamento de Física, Facultad de Ciencias Exactas, Universidad Nacional de La Plata, Calle 47 y 115, 1900 La Plata, Buenos Aires, Argentina

E-mail: danielaperez@iar-conicet.gov.ar

**Abstract.** There are several solutions of Einstein field equations that describe an inhomogeneity in an expanding universe. Among these solutions, the McVittie metric and its generalizations have been investigated through decades, though a full understanding of them is still lacking. In this note, we explore the trajectories of photons and massive particles in generalized McVittie spacetimes. In the case of massless particles, we show that no circular orbits are possible for those models that admit cosmological singularities. We also analyze the trajectory of particles for a specific generalized McVittie spacetime that is conformal to the Schwarzschild metric. By integrating the equations of motion in the Newtonian approximation, we show that particles behave in quite distinctive ways in different cosmological black hole solutions. We conclude that the analysis of the geodetic motion in inhomogeneous expanding metrics can help to discriminate those solutions that represent real cosmological black holes in the universe.

## 1. Introduction

McVittie [1] was the first to explore how the cosmic expansion affects the dynamics of local systems. He obtained an exact solution of Einstein field equations that describes a central inhomogeneity embedded in a Friedmann-Lemaître-Robertson-Walker (FLRW) cosmological background. Depending on the choice of the scale factor, a rich variety of global structures may appear. In particular, for a $\Lambda$ Cold Dark Matter ($\Lambda$CDM) model background, the solution represents a dynamical black hole [2, 3].

The McVittie metric was generalized by Faraoni and Jacques [4]; such solutions were later carefully analyzed by Carrera and Giulini [5]. These cosmological spacetimes, however, have not yet been completely characterized, let alone the associated geodetic motion. In this note, we aim to explore the trajectories of photons and massive particles for these families of metrics in order to shed some light on their characterization and interpretation.

The results of these investigations have direct astrophysical implications as shown, for instance, by Nandra and collaborators [6]. The existence of a cosmological force that manifests at large scale can introduce modifications on astrophysical structures, set a cut off on the size of galaxies and clusters of galaxies, as well as inhibit accretion processes. In a universe that is undergoing an epoch of accelerated expansion, these effects will be more drastic as time goes by, and thus, should be taken into account when modeling the future evolution of the universe.

The paper is structured as follows: in Section 2 we introduce the properties of generalized McVittie spacetimes. Then, we derive the geodesics equations and show that for a class of scale factors circular photon orbits are not possible. Next, in Section 3, we explore the conformal Schwarzschild spacetime, which is a particular case of the generalized McVittie's. After computing the geodesic equations, we obtain the corresponding Newtonian approximation and study the trajectory of particles for two different scale factors and a pair of initial conditions. In Section 4, we integrate the geodesic equations in McVittie spacetime and analyze the behavior of massive particles through cosmic time. We close the note with some conclusions.

## 2. Generalized McVittie spacetimes

We can reach McVittie metric starting with the Schwarzschild metric written in isotropic coordinates, adding a conformal factor $a(t)^2$ to the spatial part, and leaving the mass parameter $m$ to be a function of time, $m(t)$. In the case of an asymptotically spatially flat FLRW metric, i.e. $k = 0$, the metric ansatz reads:

$$ds^2 = -\left(\frac{1 - m(t)/2\tilde{r}}{1 + m(t)/2\tilde{r}}\right)^2 dt^2 + \left(1 + \frac{m(t)}{2\tilde{r}}\right)^4 a(t)^2 \left[d\tilde{r}^2 + \tilde{r}^2 d\Omega^2\right]. \tag{1}$$

The next step in McVittie's proposal regards the assumptions about the matter content.

- The matter is a perfect fluid with density $\rho$ and isotropic pressure $p$. The energy-momentum tensor is given by [5]:

$$\mathbf{T} = \rho \tilde{\mathbf{u}} \otimes \tilde{\mathbf{u}} + p\left(\tilde{\mathbf{u}} \otimes \tilde{\mathbf{u}} - \mathbf{g}\right), \tag{2}$$

where $\tilde{\mathbf{u}} := \mathbf{g}(\mathbf{u}, .)$ is the 1-form metric dual to the vector $\mathbf{u}$ that represents the four-velocity of the fluid. Notice that no equation of state is assumed.

- The four-velocity of the fluid is

$$\mathbf{u} = \mathbf{e}_0, \tag{3}$$

that is, the fluid has zero velocity with respect to the chosen reference frame. Here, $\mathbf{e}_0$ is the 0-component of the orthormal tetrad $\{\mathbf{e}_\mu\}_{\mu\in\{0,..,3\}}$ of metric (1) defined by $\mathbf{e}_\mu := \|\partial/\partial x^\mu\|^{-1}\partial/\partial x^\mu$.

Finally, Einstein field equations relate in a non-linear way the geometry of spacetime and its matter content. When solving Einstein equations, a system of 3 equations for four unknown functions $m(t)$, $a(t)$, $\rho(t,\tilde{r})$, and $p(t,\tilde{r})$ is obtained. One of these relations is:

$$\frac{d}{dt}(am) = 0, \tag{4}$$

that is directly integrated to yield

$$m(t) = \frac{m_0}{a(t)}, \tag{5}$$

being $m_0$ a constant interpreted as the mass of the central body. If we replace relation (5) into the metric (1), we obtain:

$$ds^2 = -\left(\frac{1 - m_0/2a(t)\tilde{r}}{1 + m_0/2a(t)\tilde{r}}\right)^2 dt^2 + \left(1 + \frac{m_0}{2a(t)\tilde{r}}\right)^4 a(t)^2\left(d\tilde{r}^2 + \tilde{r}^2 d\Omega^2\right). \tag{6}$$

Formula (6) is the McVittie metric, derived using the ansatz (1) and imposing conditions (2) and (3). In what follows, we call generalizations of the McVittie spacetime to metrics in which ansatz (1) is still assumed but the conditions for the matter content are not imposed.

In general, the metric (1) can be casted in areal radius coordinate‡ $R$ [8]:

$$ds^2{}_{\text{McV}} = -\left(f^2 - g^2\right)dt^2 + f^{-2}dR^2 - 2\frac{g}{f}dt\,dR + R^2 d\Omega^2, \tag{8}$$

where

$$f = \sqrt{1 - 2m(t)/R}, \tag{9}$$

$$g = R\left[H + M\left(f^{-1} - 1\right)\right], \tag{10}$$

being $M := \dot{m}(t)/m(t)$ and $H = \dot{a}(t)/a(t)$.

The causal structure of the generalized McVittie spacetime was investigated by Maciel and coworkers [8]. They found that if the spacetime background has a singularity in the past, a positive cosmological constant is assumed, and the central mass $m(t)$ has an upper limit, then depending on the asymptotic behavior of the functions $m(t)$ and

‡ The areal radius coordinate $R$ is defined by $R := \sqrt{\mathcal{A}/4\pi}$, where $\mathcal{A}$ is the area of the 2-sphere of symmetry, and where

$$d\Omega_{(2)}{}^2 := d\theta^2 + \sin\theta^2 d\phi^2, \tag{7}$$

is the line element on the unit 2-sphere [7].

$H(t)$, the spacetime contains a black hole, a pair of black and white holes, or a single white hole.

As expected, if the energy-momentum tensor is given, under certain conditions, the Schwarzschild-de Sitter and the McVittie metric are recovered [5], as shown schematically in Figure 1. In all other cases not considered in Figure 1, Einstein equations result in an under-determined system with four equations for the six functions $a$, $m$, $\rho$, $p$, and $\chi$. A possible way to arrive to a particular solution is by specifying the functions $a(t)$ and $m(t)$, and then solving Einstein equations for $q$, $\rho$, $p$, and $\chi$. Special care should be taken when proceeding in this fashion, as emphasized by Carrera and Giulini [5], to obtain physical relations between $\rho$ and $p$.

Having presented some features of this metric, we proceed now to compute the particle geodesic equations in the next subsection.

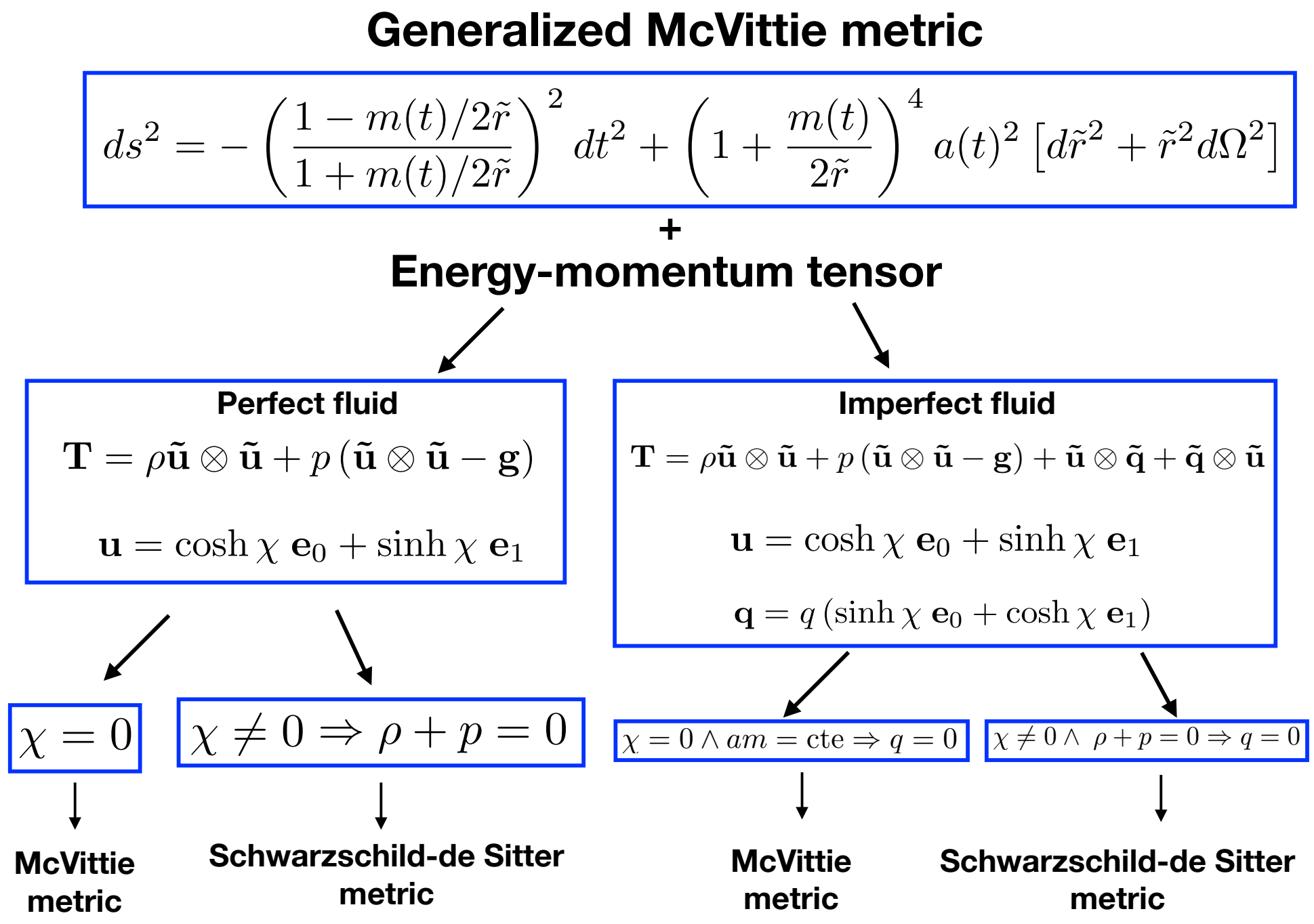

**Figure 1.** The Schwarzschild-de Sitter and McVittie metric can be recovered from the generalized McVittie metric under certain conditions. Here, $\rho$ and $p$ stand for the density and pressure of the fluid, $\mathbf{u}$ is the spherically symmetric four velocity of the fluid in an orthonormal basis; $\chi$ is the rapidity of $\mathbf{u}$ with respect to $\mathbf{e}_0$; $\mathbf{q}$ is a spatial vector field that represents the current density of heat.

*2.1. Particle geodesics in generalized McVittie spacetimes*

If we consider geodesic paths on a plane ($\theta = \pi/2$), from the spherical symmetry we obtain a conserved angular momentum:

$$\dot{\phi} = \frac{L}{R^2}, \quad \dot{\theta} = 0, \tag{11}$$

while the condition $g_{\mu\nu}\dot{x}^\mu\dot{x}^\nu = \epsilon$ yields

$$\left(-f^2 + g^2\right)\dot{t}^2 - 2\frac{g}{R}\dot{t}\dot{R} + f^{-2}\dot{R}^2 + \frac{L^2}{R^2} = \epsilon. \tag{12}$$

Here, $L$ is the angular momentum per unit mass, $\epsilon = -1$ for massive particles, and $\epsilon = 0$ for photons.

Using (11), the radial component of the geodesic equations can be written as:

$$\begin{aligned}\ddot{R} = {} & \frac{L^2}{R^3}\left(f^2 - g^2\right) + \dot{R}^2\frac{1}{f^4}\left[f^3 f' + g\left(\dot{f} - f^2 g'\right)\right] \\ & - \dot{t}^2\left[-f\dot{g} + f^3 f' - f g^2 f' + g^3\left(-\frac{\dot{f}}{f} + g'\right) + g\left(2\dot{f} - f^2 g'\right)\right] \\ & - 2\frac{\dot{t}\dot{R}}{f^3}\left[g^2\dot{f} + f^3 g f' - f^2\left(\dot{f} + g^2 g'\right)\right],\end{aligned} \tag{13}$$

where $f' = \partial f/\partial R$, $\dot{f} = \partial f/\partial t$, $g' = \partial g/\partial R$, and $\dot{g} = \partial g/\partial t$. From the latter equation, it is clear that solving for the orbit of a massive test particle around the central inhomogeneity is extremely difficult even for simple cases of $m(t)$ and $a(t)$.

In the case of photons, we can explore what are the conditions for the existence of circular orbits. In Expression (12) we impose $\ddot{R} = \dot{R} = 0$ and $\epsilon = 0$. This yields:

$$\frac{L^2}{R^2} = \left(f^2 - g^2\right)\dot{t}^2. \tag{14}$$

Notice that $\chi := f^2 - g^2 \equiv g^{\mu\nu}\nabla_\mu r \nabla_\nu R$ defines the trapped and untrapped regions of a spherically symmetric spacetime. If $R = R_\mathrm{c}$ is the radius of the photon's circular orbit, then in the regular region of the spacetime $\chi(t, R_\mathrm{c}) > 0$ for all t along the orbit:

$$\chi(t, R_\mathrm{c}) = f^2 - g^2 = 1 - \frac{2\,m(t)}{R_\mathrm{c}} - {R_\mathrm{c}}^2\left[H(t) + \frac{\dot{m}(t)}{m(t)}\left(\frac{1}{\sqrt{1 - \frac{2\,m(t)}{R_\mathrm{c}}}} - 1\right)\right]^2 > 0. \tag{15}$$

We observe that for spacetimes singular at some time in the past ($a(t_*) = 0$ for $t_* < t_0$, being $t_0$ the age of the universe) photon orbits are not possible. In this case, $\lim_{t\to t_0^+} H(t) \to +\infty$, so $\chi(t, R_\mathrm{c})$ takes negative values at early times. This result is a generalization of the one derived by Nolan [9] in McVittie spacetime.

In the next section, we specify the function $m(t)$ and use the geodesic equation (13) to investigate a particular case of the generalized McVittie metric: the conformal Schwarzschild spacetime.

## 3. Conformal Schwarzschild spacetime

### 3.1. Geodesic equations

A possible way to obtain a cosmological black hole metric is applying a conformal transformation to the whole spacetime and not just to a spatial hypersurface. We can write a family of metrics of the form:

$$\mathbf{g}_{\mathrm{cS}} := \Omega^2 \mathbf{g}_{\mathrm{Schw}}. \tag{16}$$

In Schwarzschild coordinates, this is given by:

$$\mathbf{g}_{\mathrm{cS}} = \Omega(r',t')^2 \Big[ -\Big(1 - \frac{2m_0}{r'}\Big)\mathbf{d}t'^2 + \Big(1 - \frac{2m_0}{r'}\Big)^{-1}\mathbf{d}r'^2 + r'^2 \mathbf{d}\Omega^2 \Big]. \tag{17}$$

If we choose $\Omega(r',t') \equiv \Omega(t')$ then we obtain the so-called Thakurta metric. The Thakurta metric [10] is a class of solutions of Einstein field equations that arises from a conformal transformation of the Kerr spacetime, in which the conformal factor only depends on the Boyer-Lindquist time coordinate. A special type of solutions is derived by setting the angular momentum equal to zero. The properties of such spacetime geometry were recently investigated by Mello and collaborators [11]. The metric in terms of the areal radius coordinate $R = a(t)r$ takes the form:

$$ds^2 = -\left(1 - \frac{2\tilde{M}(t)}{R} - \frac{H^2(t)R^2}{1 - \frac{2\tilde{M}(t)}{R}}\right)dt^2 + \frac{dR^2}{1 - \frac{2\tilde{M}(t)}{R}} - \frac{2\ H(t)R\ dt\ dR}{1 - \frac{2\tilde{M}(t)}{R}} + R^2 d\Omega^2. \tag{18}$$

The function $\tilde{M}(t)$ is defined as $\tilde{M}(t) = m\ a(t)$, where $a(t)$ is the scale factor, and $m$ is a constant.

Notice that the Schwarzschild metric is an Einstein space, that is:

$$R_{\alpha\beta} = \chi\ g_{\alpha\beta}, \tag{19}$$

where $\chi$ is a constant. However, when applying the conformal transformation $\Omega(\eta) = a^2(\eta)$, being $\eta$ a conformal time coordinate related to the cosmological time coordinate $t$ by $dt = a(\eta)d\eta$ [11], the conformal Schwarzschild spacetime is no longer an Einstein space, that is, metric (18) does not satisfy Eq. (19). This is what one would expect of a spacetime that is asymptotically FLRW, i.e., transforming a vacuum solution such as Schwarzschild breaks the invariance of Eq. (19).

Mello et al. [11] found that if $a(t)$ is a bounded function, the metric (18) represents a cosmological black hole. For instance, this occurs in the model with $a(t) = (\tanh t/t_0)^{2/3}$ that reproduces a dust-dominated initial era with a small transition to vacuum.

Conversely, for unbounded scale factors, metric (18) describes an inhomogeneous expanding universe [11]. A typical example of an unbounded scale factor is that of the ΛCDM model:

$$a(t) = \left[\frac{(1 - \Omega_{\Lambda,0})}{\Omega_{\Lambda,0}}\left(\sinh\left(\frac{3}{2}H_0\sqrt{\Omega_{\Lambda,0}}\ t\right)\right)^2\right]^{1/3}. \tag{20}$$

We adopt here the values $H_0 = 2.27 \times 10^{-18}\text{s}^{-1} \approx 70$ km/s Mpc, and $\Omega_{\Lambda,0} = 0.7$ for the Hubble factor and the cosmological constant density parameter, respectively.

We use the result of the last section to calculate the function $\chi(t, R)$ (see Equation (15)) for these two scale factors. The corresponding plots are displayed in Figure 2. We see that $\chi(t, R) < 0$ for some $t, R > 0$. Hence, we conclude that in the conformal Schwarzschild spacetime for bounded and unbounded scale factors with cosmological singularities, circular photon orbits do not exist.

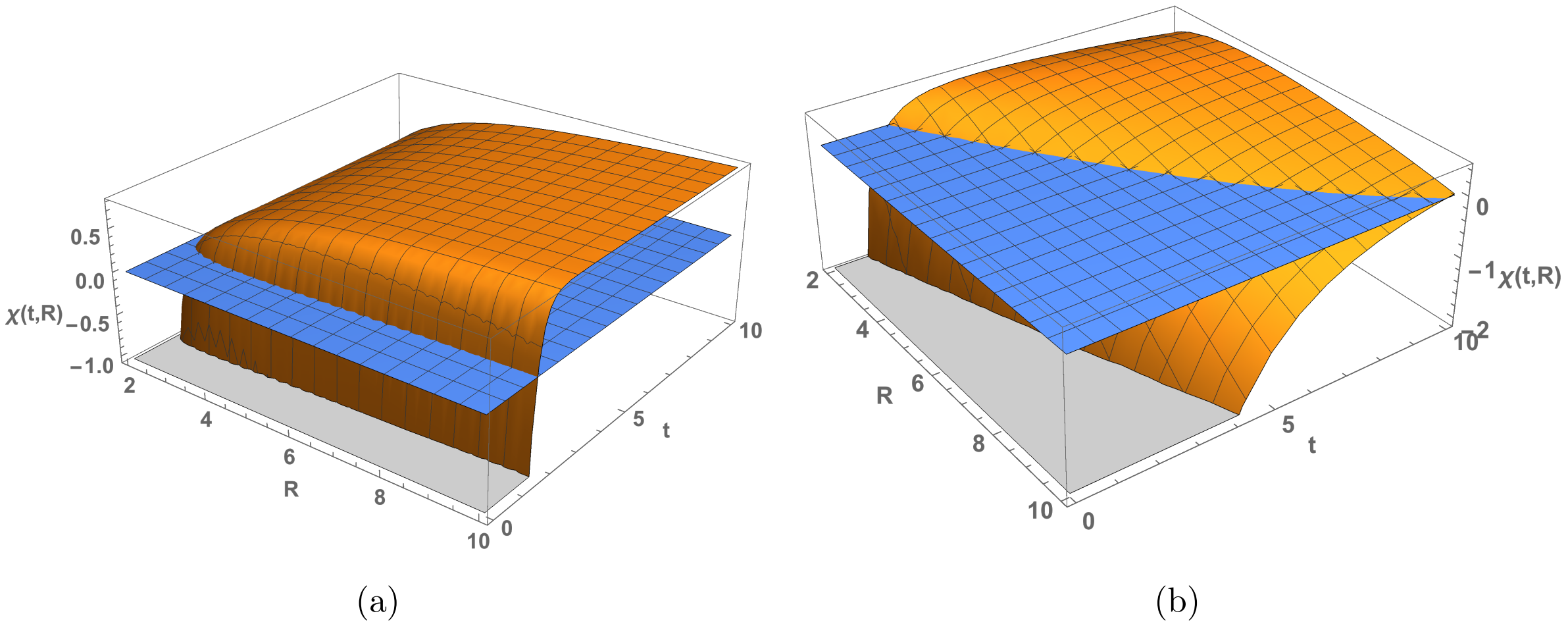

**Figure 2.** Plot of the function $\chi(t, R)$ in conformal Schwarzschild spacetime for a) scale factor $a(t) = (\tanh t/t_0)^{2/3}$; b) scale factor for the $\Lambda$CDM model. The blue horizontal plane represents $\chi(t, R) = 0$. We have set $G = c = 1$, and also $m = 1$.

The geodesic equations for a massive test particle can be derived from Equation (13) setting

$$f = \sqrt{1 - \frac{2ma(t)}{R}}, \quad g = RH(t) \left( \sqrt{1 - \frac{2ma(t)}{R}} \right)^{-1}. \tag{21}$$

After some algebraic manipulation, we obtain:

$$\ddot{R} = \frac{\dot{R}^2}{\left(1 - \frac{2\,m\,a(t)}{R}\right)^2} \left(-\alpha_1\,\delta + \gamma\right) + \frac{\gamma}{\left(1 - \frac{2\,m\,a(t)}{R}\right)} \left(1 + \frac{L^2}{R^2}\right) + \frac{L^2}{R^3}\alpha_1 - \frac{2\,\dot{R}\,\dot{t}}{\left(1 - \frac{2\,m\,a(t)}{R}\right)^2}\omega, \tag{22}$$

where,

$$\omega = \frac{m\, a'(t)\alpha_1}{R} + \frac{H(t)^2\, R^2\alpha_2}{\alpha_1} + \frac{H(t)\, R\, \left[RH'(t) + 2m\left(-aH'(t) + a'(t)\, H(t)\right)\right]}{\left(1 - \frac{2\, m\, a(t)}{R}\right)}, \tag{23}$$

$$\gamma = \frac{H(t)\, R\, \alpha_2}{\alpha_1} + \frac{RH'(t) + 2m\left(-aH'(t) + a'(t)\, H(t)\right)}{\left(1 - \frac{2\, m\, a(t)}{R}\right)} + \left(1 - \frac{2\, m\, a(t)}{R}\right)\beta, \tag{24}$$

$$\beta = \frac{-mH(t)^2\, a(t)}{\left(1 - \frac{2\, m\, a(t)}{R}\right)^2} + \frac{H(t)^2\, R}{\left(1 - \frac{2\, m\, a(t)}{R}\right)} - \frac{m\, a(t)}{R^2}, \tag{25}$$

$$\delta = -\frac{H(t)\, \left[H(t)\left(R - 4\, m\, a(t)\right) + m\, a'(t)\right]}{\alpha_1\, \left(1 - \frac{2\, m\, a(t)}{R}\right)} - \frac{m\, a(t)}{R^2}, \tag{26}$$

$$\alpha_1 = \left(1 - \frac{2\, m\, a(t)}{R} - \frac{H^2(t)R^2}{1 - \frac{2\, m\, a(t)}{R}}\right), \tag{27}$$

$$\alpha_2 = \frac{m\, a'(t)}{R} + \frac{H(t)\, R}{\left(1 - \frac{2\, m\, a(t)}{R}\right)}\left[\frac{m\, H(t)\, a'(t)}{\left(1 - \frac{2\, m\, a(t)}{R}\right)} + R\, H'(t)\right]. \tag{28}$$

We are interested in exploring the trajectories of particles in the region where both the influence of the cosmological spacetime and the local mass are important. For this task, instead of solving the extremely complex Equation (22), we resort to the Newtonian approximation which is valid in the region of interest. This approach was successfully employed in the works of Carrera and Giulini [5] and also by Nandra and collaborators [6]. The latter authors analyzed the effect of an expanding universe on massive objects. In particular, they derived the Newtonian approximation of the equation of motion for a test particle in flat McVittie spacetime, and calculated the trajectories of particles under such approximation.

In what follows, we derive the Newtonian limit of (22) and analyze the behavior of massive particles for two different scale factors.

### *3.2. Newtonian limit in conformal Schwarzschild spacetime*

In the weak field approximation, we assume:

$$\frac{m\, a(t)}{R} \ll 1, \quad R\, H(t) \ll 1. \tag{29}$$

Under this approximation, the line element (18) takes the form:

$$ds^2 = -\left(1 - \frac{2ma(t)}{R}\right)dt^2 + \left(1 + \frac{2ma(t)}{R}\right)\, dR^2 - 2\, H(t)R\, dt\, dR + R^2 d\Omega^2. \tag{30}$$

We also take the low-velocity limit $\dot{t} \approx 1$, $\dot{R} \approx 0$. If we expand Equation (22) in $m\ a(t)/R$ and $R\ H(t)$ up to first order, and also use the formula of the angular momentum $L$ in the Newtonian limit, after some algebra we finally get:

$$\frac{d^2R}{dt^2} \approx -q(t)H(t)^2R - \frac{m\,a(t)}{R^2} + \frac{L^2}{R^3}, \tag{31}$$

where $q(t) = -H'(t)/H(t)^2 - 1$ is the deceleration parameter. If we compare this expression with the one obtained by Nandra et al. (see Equation (62) in [6]) for the spatially flat McVittie spacetime, both are very similar and only differ in the mass term that is here multiplied by the scale factor.

From Equation (31), if $q(t) < 0$ then the direction of the cosmological force is opposite to the gravitational attraction caused by the central mass. For a particle moving radially ($L = 0$), there is a radius $R_0$, which is a function of the cosmic time, such that the total radial force becomes zero:

$$R_0(t) = \left[-\frac{m\,a(t)}{q(t)\;H(t)^2}\right]^{1/3}. \tag{32}$$

The radius $R_0$ marks the limit between an inner zone ($R < R_0$) and an outer zone ($R > R_0$), where the gravitational force of the central mass or the cosmological force dominates. This occurs, for instance, in the ΛCDM model. The plot of $R_0(t)$ as a function of cosmic time is shown in Figure 3. For comparison, we include in the same figure the plot of the corresponding function in McVittie spacetime (see Equation 59 in Nandra et al. [6]). For both metrics the value of $R_0$ coincides in the present time $t_0$. Contrary to the McVittie metric, when $t > t_0$ the function $R_0$ in conformal Schwarzschild increases. The presence of the scale factor in $M(t)$ is the cause of the growth of the region where gravitational attraction dominates.

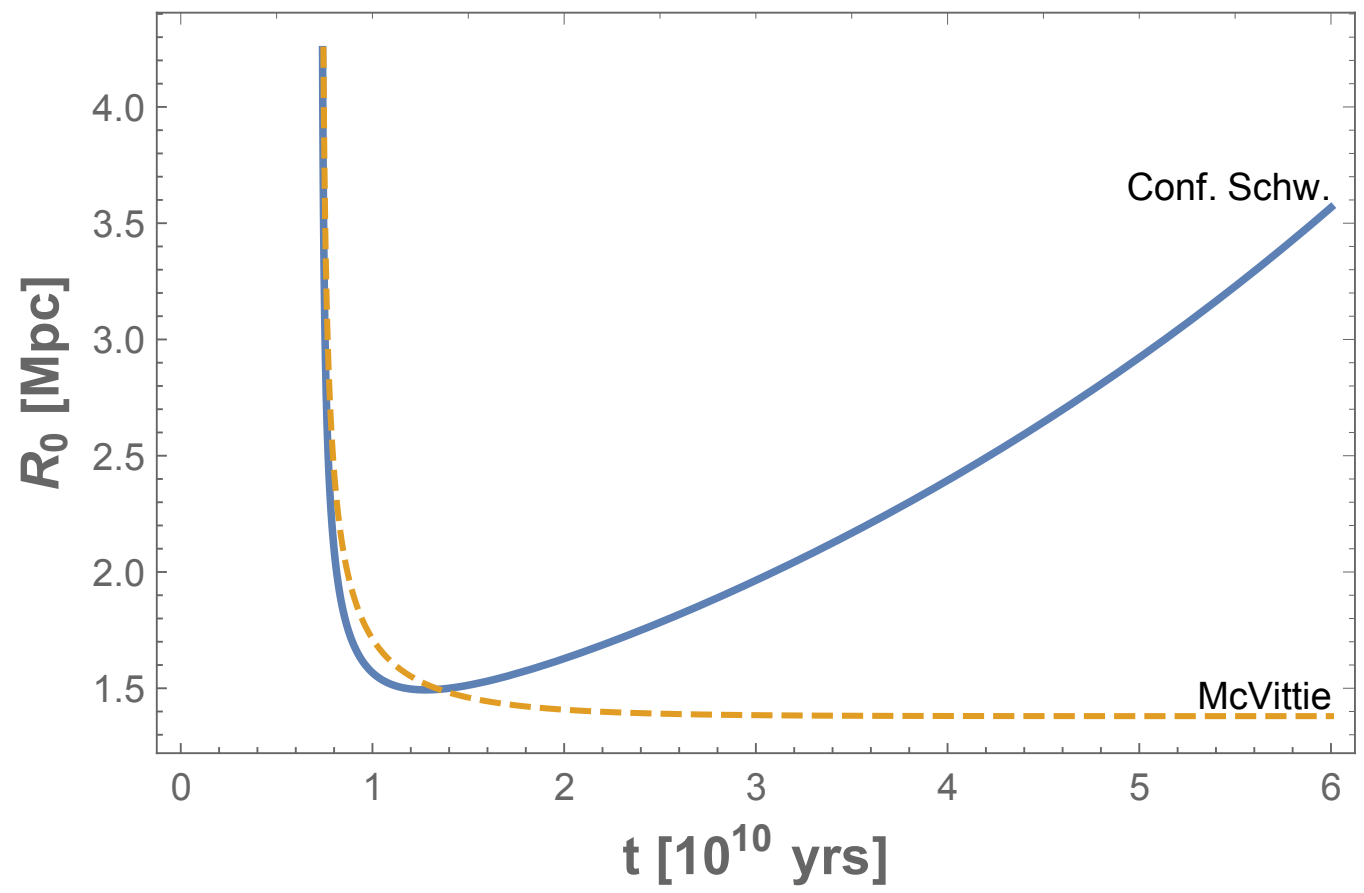

**Figure 3.** Plot of Equation (32) as a function of the cosmic time in the ΛCDM model (scale factor given by Eq. (20)). We also plot the radius at which the total radial force becomes zero in McVittie spacetime.

For bounded scale factors, as in the model with $a(t) = (\tanh t/t_0)^{2/3}$, $q(t) \geq 0$ for $t \geq 0$ and Formula (32) gives a complex number: there is no radius at which the total force vanishes; both the cosmological force and the gravitational force of the mass $m$ are directed inwards. Thus, the conformal Schwarzschild black hole metric has no radial cutoff for the existence of bounded orbits.

### *3.3. Numerical solution of the Newtonian approximation*

We now proceed to numerically solve§ Equation (31) for the scale factor of the $\Lambda$CDM model and $a(t) = (\tanh t/t_0)^{2/3}$. We choose two sets of initial conditions:

(i) A test particle is initially in a circular orbit of radius $R_\mathrm{i}$ at cosmic time $t_\mathrm{i}$, that is, $R(t_\mathrm{i}) = R_\mathrm{i}$, $dR/dt = 0$, $d^2R/dt^2 = 0$ at $t = t_\mathrm{i}$. Under these conditions, the specific angular momentum of the particle can be calculated from (31):

$$L^2 = G\ m\ a(t_\mathrm{i})R_\mathrm{i} + q(t_\mathrm{i})H(t_\mathrm{i})^2{R_\mathrm{i}}^4. \tag{33}$$

(ii) The particle is released with specific angular momentum $L = 0$ at a radius $R_\mathrm{i}$ at cosmic time $t_\mathrm{i}$, and $dR/dt = 0$, at $t = t_\mathrm{i}$.

We choose for the mass of the central object $M = 10^8\ M_\odot$.

- First set of initial conditions, $\Lambda$CDM scale factor.
  We show in Figure 4(a) the results of the numerical integration. For comparison, we also solve Equation (31) for the flat McVittie spacetime (see Figure 4(b)). For the two cases, the particle is initially in a circular orbit of radius $R_\mathrm{i} = 25$ Kpc at the present epoch.
  There are notable differences in the trajectory of the particle between both spacetimes. In Figure 4(a), we see that the radius initially decays, then it starts growing till a certain point, next it decays again, and so on. The amplitude of this periodic behavior gets smaller, and so does the mean radius of the particle orbit as cosmic time increases. In Figure 4(b), the radius of the particle oscillates around a certain value that stays constant through cosmic time. A possible explanation for the dissimilarities between these metrics is that in the conformal Schwarzschild spacetime the presence of the scale factor in the mass term increases the gravitational attraction exerted on a particle as time goes by; as a consequence, the particle's orbit shrinks and finally falls into the central object.
- First set of initial conditions, $a(t) = (\tanh t/t_0)^{2/3}$ scale factor.
  The behavior of the particle is practically the same as in the $\Lambda$CDM model, as depicted in Figure 5(a). We see from Equation (31) that the decreasing oscillatory radius of the particle is due to the interplay between the two negative terms (central

§ The equation of motion (31) was solved numerically using the computer program Wolfram Mathematica version number 11.2.0. More specifically, Eq. (31) was integrated with the program's function "NDSolve". The method of integration was chosen automatically by the function in order to maximize the accuracy of the result. For more details on Wolfran Mathematica and the function "NDSolve", the reader is referred to `http://www.wolfram.com/mathematica/` and `https://reference.wolfram.com/language/ref/NDSolve.html`.

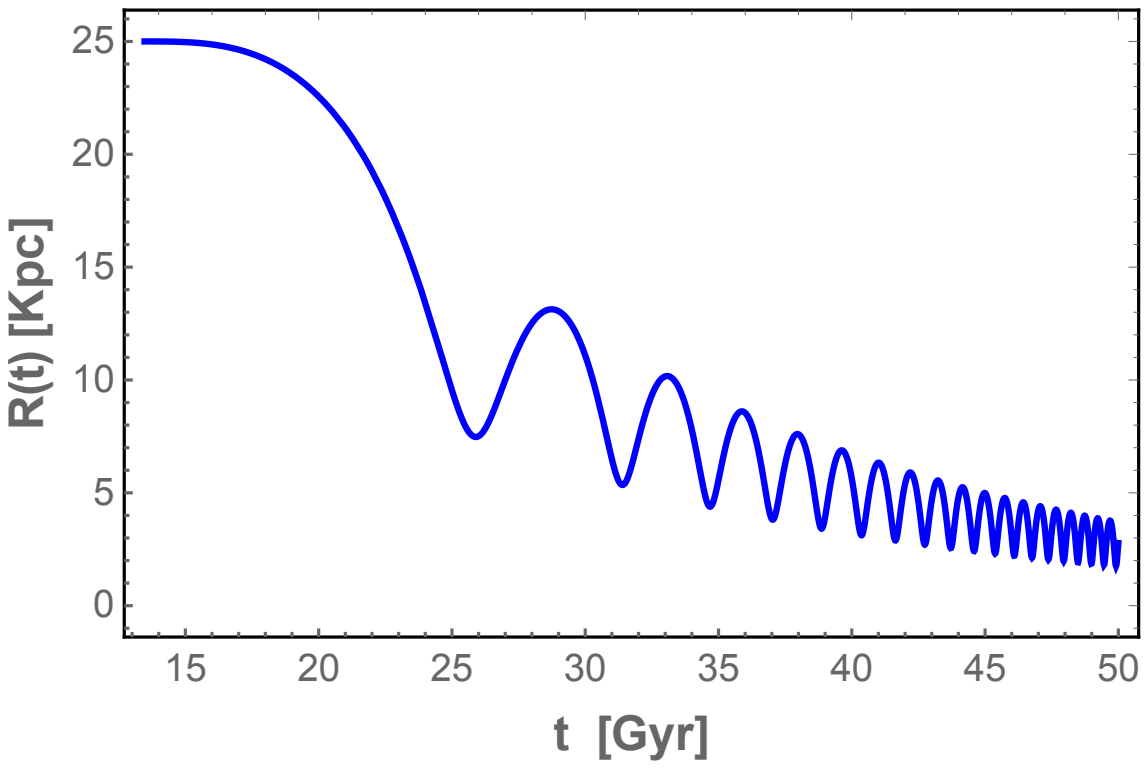

(a)

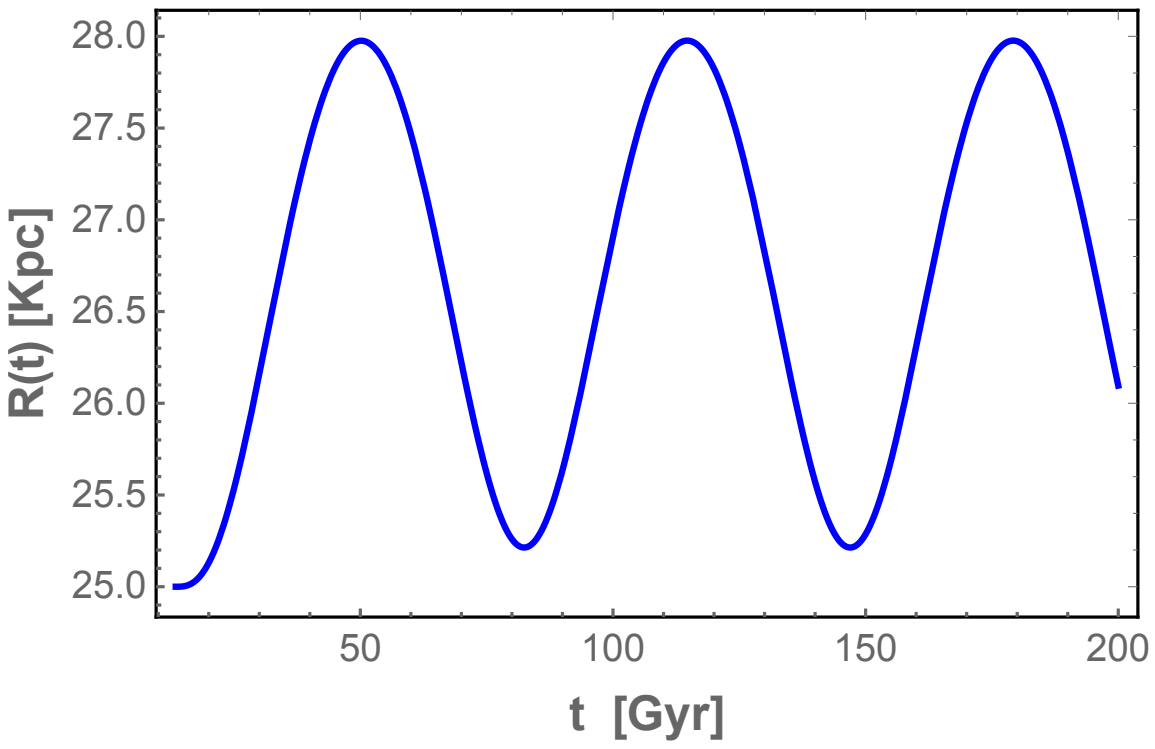

(b)

**Figure 4.** Radial coordinate of a test particle as a function of the cosmic time that is launched from a circular orbit of radius $R_\mathrm{i} = 25$ Kpc at $t_\mathrm{i} = t_0 = 13.47$ Gyr in the Newtonian approximation for: a) conformal Schwarzschild using the ΛCDM model scale factor, and b) flat McVittie spacetime.

mass and cosmological force) and the angular momentum term that is always positive.

- Second set of initial conditions, ΛCDM scale factor.
  The trajectory of the particle highly depends on its initial radius (see Section 3.2). If $0 < R_\mathrm{i} < 55$ Kpc, the particle falls directly into the central object. For $55$ Kpc $\leq R_\mathrm{i} \leq 65$ Kpc, the particle is initially driven by the cosmological expansion until the scale factor in the mass term begins to dominate and it is attracted towards the central mass (see Figure 5(b)) For $R_\mathrm{i} > 65$ Kpc, the particle moves with the cosmological expansion.
- Second set of initial conditions, $a(t) = (\tanh t/t_0)^{2/3}$ scale factor.
  The particle always ends up in the black hole independently of the initial radius of the orbit.

We close this note computing the particle's geodesics in flat McVittie spacetime.

## 4. Geodesics in flat McVittie spacetime

The geodesics equations can be obtained from formulae (12) and (13) setting: $f = \sqrt{1 - 2m_0/r}$, and $g = rH$. It is useful to express these equations in terms of the Hubble factor. In the ΛCDM model:

$$H(t) = H_\infty \coth(t/t_\Lambda), \quad w(H) := \frac{dH}{dt} = \frac{3}{2}(H_\infty^2 - H^2), \tag{34}$$

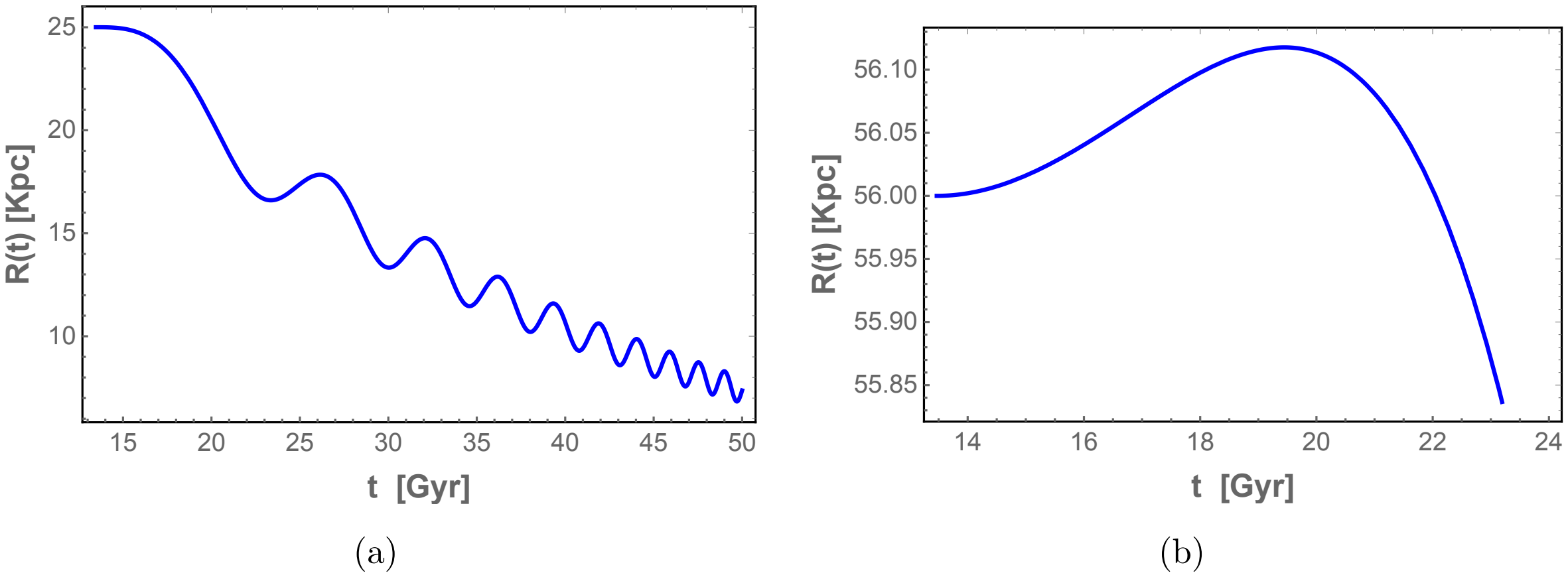

**Figure 5.** Plot of the numerical solution of Equation (31) as a function of the cosmic time $t$ for the newtonian approximation of the conformal Schwarzschild spacetime with: a) first set of initial conditions and scale factor $a(t) = (\tanh t/t_0)^{2/3}$; b) second set of initial conditions in the $\Lambda$CDM model.

where $H_\infty = H_0\sqrt{\Omega_\Lambda}$ and $t_\Lambda = (2/3)/H_\infty$. Under the change of coordinates $t \to H$, the geodesic equations take the form:

$$\frac{dr}{dH} = \frac{p}{\dot{t}w}, \tag{35}$$

$$\frac{dp}{dH} = \frac{\dot{p}}{\dot{t}w} = \frac{1}{\dot{t}w}\left[r\sqrt{1-\frac{2m_0}{r}}wu^2 + \left(1-\frac{3m_0}{r}\right)\frac{l^2}{r^3} - \left(\frac{m_0}{r^2} - H^2 r\right)\right]. \tag{36}$$

In this spacetime, for almost all scale factors $a(t)$ there are not circular orbits; however, bounded asymptotic circular orbits for the $\Lambda$CDM model are possible, as Nolan demonstrated in a series of theorems [9]. We use the latter result to explore how these orbits behave through the cosmic history of the universe.

The initial conditions chosen are $p(H_\infty) = 0$ and $r(H_\infty) = R_C(l,\sigma)$, i.e. the orbits are asymptotic to circular orbits in Schwarzschild-de Sitter spacetime with angular momentum $l$ and mass $\sigma$. We integrate Equations (35) and (36) for $H_\infty \leq H \leq 15\ H_\infty$, where $H_\infty = 1.98 \times 10^{-4}$ Mpc$^{-1}$ in geometrized units ($G = c = 1$). The corresponding time interval is $7.32 \times 10^8$ yrs $\leq t < \infty$, that is, from the time the first stars and galaxies formed till the spacetime geometry of the universe is fully de Sitter. The results for different values of $R_C$‖ are shown in Figure 6.

From Figures 6(a) and 6(b) we see that our results confirm Nolan's theorems. In flat McVittie spacetime for the $\Lambda$CDM model, the orbits of massive particles are bounded. The radius of such orbits increases with time. Bounded systems are more compact in the early universe.

In order to check our calculations, we integrate again Equations (35) and (36), but taking for the initial conditions $p(2H_\infty) = 0$ and $r(2H_\infty) = R_C(l,\sigma)$: we assume the existence of circular orbits whose radius corresponds to the Schwarzschild-de Sitter spacetime at a different moment of time. The result is plotted in Figure 7. Though the radius of the orbit is almost constant in time, for $H < 2H_\infty$ it starts to grow indefinitely, and the orbit becomes unbounded.

‖ Circular orbits in de Schwarzschild-de Sitter spacetime have an innermost stable circular orbit (ISCO) and also an outermost stable circle orbit (OSCO).

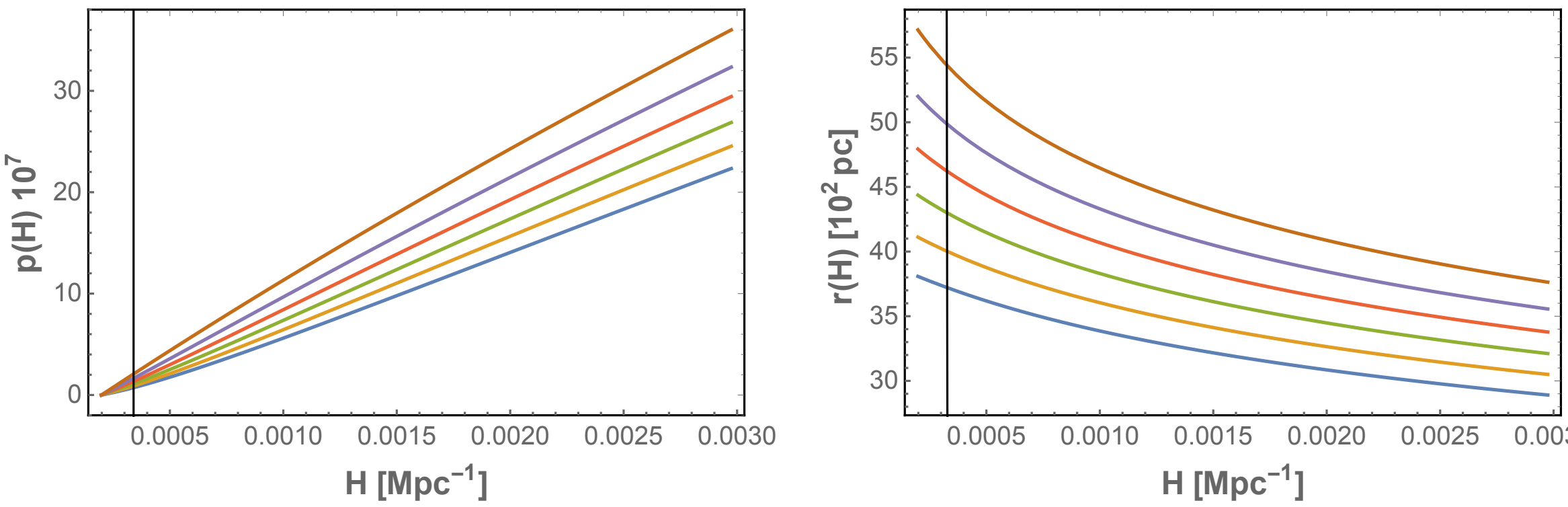

(a) (b)

**Figure 6.** Plot of the numerical solution of Equations (35) and (36) as a function of the Hubble factor $H$ in flat McVittie spacetime: a) specific linear momentum $p$ , and b) radial coordinate $r$ of a test particle. The vertical line indicates the value of $H$ at which the cosmic expansion changes direction (deceleration parameter $q = 0$). The mass of the central object is $M = 10^6\ M_\odot$, similar to what is thought was the mass of the black hole in the first quasars.

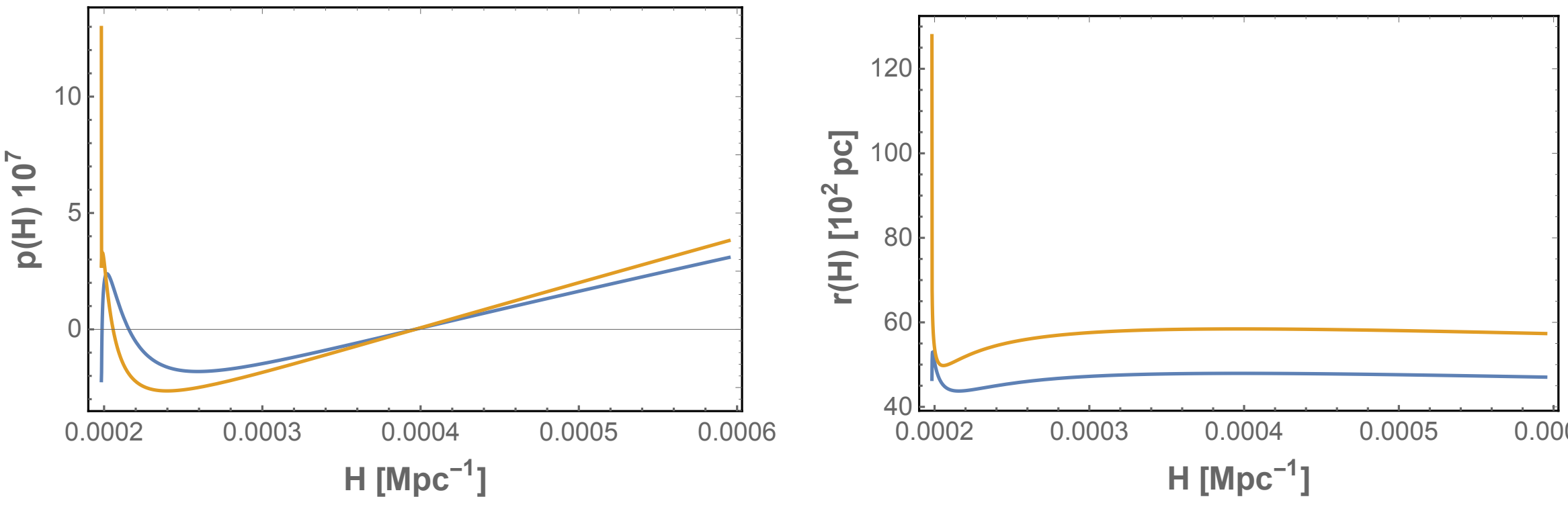

(a) (b)

**Figure 7.** Plot of the numerical solution of Equations (35) and (36) as a function of the Hubble factor $H$ in flat McVittie spacetime for the initial conditions $p(2H_\infty) = 0$ and $r(2H_\infty) = R_C(l, \sigma)$. The mass of the central object chosen is $M = 10^6\ M_\odot$.

## 5. Conclusions

We have investigated the trajectories of both massive particles and photons in inhomogeneous expanding spacetimes. Our work has been focused on a particular family of cosmological metrics dubbed generalized McVittie spacetimes which in the appropriate limit reduce to the well-studied McVittie solution. We first computed the geodesic equations. For massless particles, we found that circular orbits are not possible in those spacetime models that admit cosmological singularities.

A particular member of this class is the conformal Schwarzschild metric. For this specific spacetime, we showed that circular photon orbits do not exist. We also calculated the geodesic equations. Since such system of equations is very complex to solve, a first approach to gather a general understanding of the behavior of the particles is to study the corresponding Newtonian approximation. In the region where the approximation is valid, the particle is neither too close to the central object nor at cosmological distances, so will be equally affected by both gravitational attraction of the inhomogeneity and the cosmological expansion.

Using this approximation, we found that there is no radial cut off for the existence of bounded orbits in the case the metric represents a black hole. This is not the case if the conformal Schwarzschild spacetime represents an inhomogeneous expanding universe, or for other cosmological black hole solutions such as the flat McVittie metric in the ΛCDM model. These results have some astrophysical implications: if galaxies harbor a dominant supermassive black hole at their core, the cosmological background would play no role in setting their maximum size. On the contrary, the cosmological repulsive force would impose an upper limit on the radius of cluster of galaxies and structures whose dynamics is not dominated by a central compact object.

We integrated numerically, under the Newtonian approximation, the orbit of massive particles for two different set of initial conditions: i) particles that are initially in a circular orbit, and ii) particles that are released with specific angular momentum $L = 0$. For the first set of initial conditions, the evolution of the particle's orbit is almost identical for bounded and unbounded scale factors; on a global time scale the radius of the particle decays, though for shorter periods of time it oscillates with an ever decreasing amplitude. This behavior is completely different compared to the McVittie cosmological black hole solution. For the second set of initial conditions and a bounded scale factor, the particle always ends up in the black hole independently of its initial location. For unbounded scale factor, however, the initial radius of the particle determines whether it falls into the central object or is driven by the cosmological expansion.

We conclude the analysis of particle's orbits integrating the geodesic equations in the flat McVittie metric. We found that in the ΛCDM model, the orbits of massive particles are restricted in a range of radii through the history of the universe; bounded systems are more compact in the early universe. The results obtained confirmed some of Nolan's theorems [9].

The trajectories of particles in inhomogeneous expanding spacetimes have very specific features that distinguish them from the ones present in stationary black holes and also in cosmological models. These differences manifest among the various cosmological black hole solutions as we have shown in this note.The analysis of the geodetic motion, thus, helps not only to better characterize these dynamical metrics but also contribute to the understanding of the underlying cosmological model.

## Acknowledgments

This work was supported by grants AYA2016-76012- C3-1-P (Ministerio de Educación, Cultura y Deporte, España) and PIP 0338 (CONICET, Argentina).